Transport properties of p-type metal-oxide-semiconductor inversion layer in (110) and (111) silicon channel under uniaxial stress

Fan Jun Wei*

*Institute of Physics, Academia Sinica, 128 Sec. 2, Academia Rd., Nankang, Taipei 11529, Taiwan, ROC*

## Abstract

Valence subband and transport properties of p-type metal-oxide-semiconductor (PMOS) inversion layer in uniaxially strained (110) and (111) Si channel have been studied theoretically in this work. Equal energy lines, carrier concentration effective mass, conductivity effective mass, and mobility are calculated based on Luttinger-Kohn Hamiltonian [16]. Three compact expressions of scattering rate are derived in this paper. PMOS channel direction is chosen as 45° in x-y plane, namely, $[-1,\ \ 1,\ \ -\sqrt{2}]$ in (110) system and $[1-\sqrt{3},\ \ 1+\sqrt{3},\ \ -2]$ in (111) system and z direction is chosen to be the quantization direction. The direction of applied uniaxial stresses considered is in either parallel or perpendicular to channel. My results show mobility is very sensitive to effective mass under different stresses. The favorable mobility directions are found in (110) system with -3 GPa stress parallel to channel and +3 GPa stress perpendicular to channel.

keywords: inversion layer ; transport ; (110) Si ; (111) Si ; scattering ; uniaxial stress ; effective mass ; mobility

---

*e-mail address: alanfan101@yahoo.com

## 1. Introduction

It has been shown strain technology in complementary metal–oxide–semiconductor (CMOS) results in favorable transport properties [3-4]. Recently, researchers are trying to develop various novel structures, such as fin field-effect transistor (FinFET) or multiple gate field-effect transistor (MuGFET), for next generation devices. However, strain technology remains an effective and economical manipulation to produce higher carrier mobility for electronic devices [5]. In general, strains can be created to channel by lattice mismatch between channel and substrate or through the external mechanical stresses, or combinations of these two methods. Biaxial strains produced by lattice mismatching has been proven effective of lowering carrier effective masses but uniaxial stress applied to the channel in some specific substrate oreingtations can also create favorable transport properties in the experiments [6-13]. In this work, I theoretically calculated the equal energy lines, carrier concentration effective mass ($m_c$), conductivity effective mass ($m_\sigma$), and hole mobility ($\mu_{jj}$) under uniaxial stresses -3 ,-2, -1, 0, 1, 2, 3 GPa etc. of (110) and (111) Si channel, respectively. Channel direction is chosen as 45° direction in x-y plane, namely, $[-1,\ 1,\ -\sqrt{2}]$ in (110) system and $[1-\sqrt{3},\ 1+\sqrt{3},\ -2]$ in (111) system. The z direction is chosen to be the quantization direction. The direction of the applied stress considered is either parallel or perpendicular to channel and unaxial stresses can be can be realized by doping carbon or germanium into the opposite sides of the Si channel.

## 2. Physics Methods

### 2.1. Luttinger-Kohn Hamiltonian

On the basis of $k \cdot p$ theory, the valence energy subband structures of diamond structures under strain can be found through diagonalizing a 6×6 Luttinger-Kohn

Hamiltonian [16]. The electric field produced by the gate bias, which gives rise to the quantum confinement effects, is assumed to be a constant and quantizes the z-component of momentum $\hbar k_z$ along the z-direction. Consequently $\hbar k_z$ is expressed as $\frac{\hbar}{i}\frac{\partial}{\partial z}$. I discretize the linear potential energy into $N$ equal height steps such that the Hamiltonian matrix for the inversion layer is constructed as $H_{FEM}$ [7],

$$H_{FEM} = \begin{bmatrix} \ddots & \vdots & \vdots & \vdots & \vdots & \vdots & \\ & & & & & & \\ \cdots & D_- & D_{n-1} & D_+ & 0 & 0 & \cdots \\ \cdots & 0 & D_- & D_n & D_+ & 0 & \cdots \\ \cdots & 0 & 0 & D_- & D_{n+1} & D_+ & \cdots \\ & & & & & & \\ & \vdots & \vdots & \vdots & \vdots & \vdots & \ddots \end{bmatrix}, \qquad (1)$$

with

$$\begin{aligned} H_{LK} &= H_0 + H_1 k_z + H_2 k_z^2 \\ D_n &= H_0 + 2\frac{H_2}{(\Delta z)^2} + V_n \\ D_+ &= \frac{H_1}{2i\Delta z} - \frac{H_2}{(\Delta z)^2} \\ D_- &= -\frac{H_1}{2i\Delta z} - \frac{H_2}{(\Delta z)^2} \end{aligned},$$

where $V_n$ is the potential energy value of the $n^{th}$ step, and $\Delta z = 0.2$ nm gives the width of each step. We take the total number of step $N=50$ for the calculations. Due to the strain effect each valence subband splits into a set of two very close subbands, which correspond to spin up and down, respectively. The three sets of lowest energy valence subbands, in increasing order of energy, are labeled as the 1$^{st}$ subband, the 2$^{nd}$ subband, and the 3$^{rd}$ subband. The strain tensors used in our work are adopted from Ma *et al* [17]. So far we have all the details of Energy-Momentum dispersions of 2D carriers with strains.

## 2.2. Scattering Formulas

Instead of the conventional scattering formulas [18-22], three compact expressions of scattering rate including accoustic phonon, optic phonon, and, surface roughness scatterings have been derived in this work, which are following Harrison's [1] and Lundstrom's [2] formulas.

The $n^{th}$ acoustic phonon scattering rate in the inversion layer is derived as

$$\frac{1}{\tau_n(E)} = \frac{D_A^2 m_c k_B T}{\rho v_g^2 (2\pi)^2 \hbar^3} \times \theta\left( k_n^2(E) - \frac{2m_c \Delta}{\hbar^2} \right) \pi \int_0^\infty G^2(K_z) dK_z \,, \tag{2}$$

where $G(K_z) = \int \psi_f^*(z) e^{-iK_z z} \psi_i(z) dz$, in which $K_z$ is the acoustic phonon wave number in the z direction, and $\theta\left( k_n^2(E) - \frac{2m_c \Delta}{\hbar^2} \right)$ is the unit step function, in which $\Delta \equiv Ez_f - Ez_i$, $Ez_f$ and $Ez_i$ give the final and initial quantization energy in the z direction, respectively. $D_A$ is acoustic phonon deformation potential energy, $\rho$ is Si mass density, $m_c$ is carrier concentration effective mass, $k_B$ is Boltzmann's constant, and $v_g$ is acoustic phonon group velocity. The subscript $n$ represents the $n^{th}$ subband in the whole place of this paper.

The $n^{th}$ optic phonon scattering rate is derived as

$$\frac{1}{\tau_n(E)} = \frac{D_o^2 m_c}{\rho \omega_o (2\pi)^2 \hbar^2} \left( N_o + \frac{1}{2} \mp \frac{1}{2} \right) \times \theta\left( k_n^2(E) - \frac{2m_c \Delta}{\hbar^2} \right) \pi \int_0^\infty G^2(K_z) dK_z \,, \tag{3}$$

where $G(K_z) = \int \psi_f^*(z) e^{-iK_z z} \psi_i(z) dz$, in which $K_z$ is the acoustic phonon wave number in the z direction, and $\theta\left( k_n^2(E) - \frac{2m_c \Delta}{\hbar^2} \right)$ is the unit step function, in which $\Delta \equiv Ez_f - Ez_i \mp \hbar\omega_o$, $Ez_f$ and $Ez_i$ give the final and initial quantization energy in the z direction, respectively. $\hbar\omega_o$ gives hole optic phonon energy and $D_o$ is optic phonon deformation potential. $N_o$ is the Bose-Einstein's function, because the phonons are bosons.

The $n^{th}$ surface roughness scattering rate is derived as

$$\frac{1}{\tau_n(E)} = \frac{\pi m_c}{\hbar^3} \frac{(eF\Delta L)^2}{1+\frac{L^2 k_n^2}{2}} \int_0^\infty G^2(K_z) dK_z \,, \tag{4}$$

where $L$ and Δ are surface roughness parameters. Furthermore, Δ is root-mean-square amplitude of fluctuations and $L$ is their correlation length, which is roughly the distance between fluctuations. $F$ is electric field strength which is chosen as 1 MV/cm in all stress cases in this paper.

## 2.3. Carrier Concentration Effective Mass

Carrier concentration effective mass ($m_c$), is calculated according to ref. [7]. $m_c$ is also a parameter for calculating mobility.

$$m_{c,n} = \frac{\pi\hbar^2 \int_{E_0}^{\infty} D_n(E) f(E) dE}{\int_{E_0}^{\infty} f(E) dE} \,, \tag{5}$$

where $D_n(E)$ is density of states of holes and $f(E)$ is Boltzmann's function, because the Fermi's level is canceled [7].

## 2.4. Conductivity Effective Mass

Conductivity effective mass ($m_\sigma$), is calculated according to ref. [7].

$$m_{\sigma,j} = \frac{\int_{E_0}^{\infty} D_n(E) f(E) dE}{\int_{E_0}^{\infty} D_n(E) v_j^2(E) \left( -\frac{\partial f(E)}{\partial E} \right) dE} \,, \tag{6}$$

where $v_j$ is the $j^{th}$ component of hole group velocity, $\tau(E)$ is hole scattering time. Similar to the case of carrier concentration effective mass, Fermi's level is also canceled.

## 2.5. Mobility

Carrier mobility ($\mu_{jj}$), is also calculated according to ref. [7].

$$\mu_{jj} = \frac{e\int_{E_0}^{\infty}\sum_{n=1}^{3} D_n(E)v_j^2(E)\tau(E)\left(-\frac{\partial f(E)}{\partial E}\right)dE}{\int_{E_0}^{\infty}\sum_{n=1}^{3} D_n(E)f(E)dE}, \tag{7}$$

where $e$ is carrier charge of a hole. The results of $m_c$, $m_\sigma$, and $\mu_{jj}$, in (110) and (111) systems are all under applied electric field strength 1 MV/cm with external stresses from -3 GPa to +3 GPa either parallel or perpendicular to channel. All calculations are performed by Luttinger-Kohn parameters of Si [14-17] and scattering parameters of Si shown in TABLE I [2].

## 3. Results and Discussions

### 3.1. Equal Energy Lines

For (110) Si channel system, Fig. 1, 2, and 3 show the 25 meV equal energy lines of the three lowest energy valence subbands (only one spin state with internal strain are shown) of the inversion layer with different uniaxial stresses of -3 ,-2, -1, 0, 1, 2, 3 GPa, respectively. Fig. 4, 5, and 6 show the same thing as in Fig. 1, 2, and 3 for (111) Si channel system. The choice of 25 meV as a demonstrated energy for plotting the equal energy lines in Fig. 1 to 6 is because the thermal energy is 25 meV and is also because I follow the Fishetti's choice [18]. We can see equal energy lines in (110) Si channel are elliptic and when the external stress getting large the major axes of the ellipse lie on different directions for the 1$^{st}$ and the 2$^{nd}$ subband while the 3$^{rd}$ subband become more wrap than the other subbands. When compress stresses are added in the (110) system equal energy lines turn the major axes of the ellipse into negative degree and tensile stresses cause opposite effects to compress stresses. The symmetry of the equal energy lines in (111) system are not elliptic but hexagonal. As the external stress are increasing in the (111) system, equal energy lines are become ellipse from hexagon and the major axes lie also on the negative degree when the applied stress are compress for the 1st, 2nd and 3rd subband and tensile stresses cause the opposite effects to the major axes of the 1st, 2nd and 3rd subband.

## 3.2. Carrier Concentration Effective Mass

Fig. 7 and 8 show $m_c$ of the three lowest valence subbands as a function of external stresses in (110) and (111) Si channel, respectively. Comparing Fig. 7 and 8 we find generally speaking at a given external stress, $m_c$ for each subband in (111) system is larger than in (110) system except for the 3rd subband in the cases of large compressive and tensile stresses. The trends of the $m_c$ variations are different in (110) and (111) Si systems when external stresses are added. In (110) system, $m_c$ of the 1st and 2nd subband increase slightly when external stresses are enhanced in both compressive and tensile cases but $m_c$ of the 3rd subband increases monotonically from -3 GPa to +3 GPa. In (111) system, $m_c$ of the 1st and 3rd subband decrease when external stresses are enhanced in both compressive and tensile stresses but $m_c$ of the 2nd subband behaves oppositely. The reason of the difference is due to the 1st and 2nd subband are both heave hole like but the 3rd subband is light hole like in (110) Si channel and the 1st and 3rd subband in (111) channel are heavy hole like but the 2nd subband is light hole like.

## 3.3. Conductivity Effective Mass

Fig. 9 and 10 show conductivity effective mass ($m_\sigma$) in both (110) and (111) channels. It can be seen $m_\sigma$ of the three lowest subbands in (110) system are generally speaking smaller than (111) system, but in either (110) or (111) system, it is lowered by compressive stresses applied in the direction parallel to channel and tensile stresses applied perpendicular to channel. The channel direction is chosen as along the 45° direction in the x-y plane, namely, $[-1 \quad 1 \quad -\sqrt{2}]$ in the (110) system and $[1-\sqrt{3} \quad 1+\sqrt{3} \quad -2]$ in the (111) system.

## 3.4. Mobility

Results of the mobility ($\mu_{jj}$) in (110) and (111) channels under external electric field strength 1 MV/cm with external stresses applied either parallel or perpendicular to channel are shown in Fig. 11. The external stress conditions of $\mu_{jj}$ are all same as the conditions of $m_\sigma$. We can see $\mu_{jj}$ in (110) system is in general larger than (111) system, but in either (110) or (111) system, it is arisen by compressive stresses applied in the direction parallel to channel and tensile stresses applied perpendicular to channel. After comparing the behaviors of $m_\sigma$ and $\mu_{jj}$, my results show mobility is very sensitive to effective mass under different stresses.

## 4. Conclusions

In this work I reported theoretical analysis of valence energy subband properties including the hole mobility of p-type metal-oxide-semiconductor (PMOS) inversion layer with both compressive and tensile external stresses in both (110) and (111) Si channels. Three compact expressions of scattering rate are also derived in this paper. Channel direction is chosen as along the $45^\circ$ direction in the x-y plane, namely, $[-1 \quad 1 \quad -\sqrt{2}]$ in (110) system and $[1-\sqrt{3} \quad 1+\sqrt{3} \quad -2]$ in (111) system. Uniaxial stresses are considered in the directions both parallel and perpendicular to channel. Compressive and tensile stresses considered in this paper can be realized by doping carbon or germanium into the opposite sides of Si channel. Equal energy lines, carrier concentration effective mass ($m_c$), conductivity effective mass ($m_\sigma$), and hole mobility ($\mu_{jj}$) under uniaxial stresses -3 ,-2, -1, 0, 1, 2, 3 GPa etc. of (110) and (111) Si channels are all calculated and discussed.

## References

[1] Paul Harrison, *Quantum Wells, Wires and Dots: Theoretical and Computational Physics of Semiconductor Nanostructures*, Wiley, (2010).

[2] Mark Lundstrom, *fundamental of carrier transport*, Cambridge University Press, (2000).

[3] Minjoo L. Lee, Eugene A. Fitzgerald, Mayank T. Bulsara, Matthew T. Currie, and Anthony Lochtefeld, J. Appl. Phys. **97** (2005) 011101.

[4] Anh-Tuan Pham, Christoph Jungemann, and Bernd Meinerzhagen, *IEEE* Trans. Electron Devices **54** (2007) 2174.

[5] Y. Zhang, M. V. Fischetti, B. Sorée, W. Magnus, M. Heyns, and M. Meuris, J. Appl. Phys. **106** (2009) 083704.

[6] Agostino Pirovano, Andrea L. Lacaita, Günther Zandler, and Ralph Oberhuber, *IEEE* Trans. Electron Devices **47** (2000) 718.

[7] Shu-Tong Chang, Jun Wei Fan, Chung-Yi Lin, Ta-Chun Cho, and Ming Huang, J. Appl. Phys. **111** (2012) 033712.

[8] L. Donetti, F. Gámiz, S. Thomas, T. E. Whall, D. R. Leadley, P.-E. Hellstrom, G. Malm, and M. Ostling, J. Appl. Phys. **110** (2011) 063711.

[9] Bing-Fong Hsieh and Shu-Tong Chang, Solid-State Electronics **60** (2011) 37.

[10] S. T. Chang, Jacky Huang, Ming Tang, and C. Y. Lin, Thin Solid Films **518** (2010) S154.

[11] Guangyu Sun, Yongke Sun, Toshikazu Nishida, and Scott E. Thompson, J. Appl. Phys. **102** (2007) 084501.

[12] S. T. Chang, S. H. Liao, and C. Y. Lin, Thin Solid Films **517** (2008) 356.

[13] C. Gallon, G. Reimbold, G. Ghibaudo, R.A. Bianchi, R. Gwoziecki, Solid-State Electronics **48** (2004) 561.

[14] D. Rideau, M. Feraille, L. Ciampolini, M. Minondo, C. Tavernier, H. Jaouen, and

A. Ghetti, Phys. Rev. B **74** (2006) 195208.

[15] Yong-Hua Li, X. G. Gong, and Su-Huai Wei, Phys. Rev. B **73** (2006) 245206.

[16] Shun Lien Chuang, *Physics of Optoelectronic Devices*, Wiley, (1995).

[17] Q. M. Ma, and K. L. Wang, and J. N. Schulman, Phys. Rev. B **47** (1993) 1936.

[18] M. V. Fischetti, Z. Ren, P. M. Solomon, M. Yang, and K. Rim, J. Appl. Phys. **94** (2003) 1079.

[19] M. V. Fischetti and S. E. Laux, J. Appl. Phys. **80** (1996) 2234.

[20] Ying Fu, Kaj J. Grahn, and Magnus Willander, *IEEE* Trans. Electron Devices **41** (1994) 26.

TABLE I.

Scattering parameters of Si are shown in TABLE I [2].

| | | |
|---|---|---|
| Mass density (g/cm$^3$) | $\rho$ | 2.329 |
| Hole acoustic phonon group velocity (m/s) | $v_g$ | 9040 |
| Hole acoustic phonon deformation potential energy (eV) | $D_A$ | 5.0 |
| Hole optic phonon deformation potential (× $10^8$eV/cm) | $D_o$ | 6.00 |
| Hole optic phonon energy (eV) | $\hbar\omega_o$ | 0.063 |
| Root-mean-square amplitude of fluctuations (nm) | $A$ | 0.4 |
| Correlation length (nm) | $L$ | 3 |

## Figure Captions

Fig. 1 The 25 meV equal energy lines of the 1st valence subband at gate electric field 1 MV/cm as a function of different stresses in (110) Si channel.

Fig. 2 The 25 meV equal energy lines of the 2nd valence subband at gate electric field 1 MV/cm as a function of different stresses in (110) Si channel.

Fig. 3 The 25 meV equal energy lines of the 3rd valence subband at gate electric field 1 MV/cm as a function of different stresses in (110) Si channel.

Fig. 4 The 25 meV equal energy lines of the 1st valence subband at gate electric field 1 MV/cm as a function of different stresses in (111) Si channel.

Fig. 5 The 25 meV equal energy lines of the 2nd valence subband at gate electric field 1 MV/cm as a function of different stresses in (111) Si channel.

Fig. 6 The 25 meV equal energy lines of the 3rd valence subband at gate electric field 1 MV/cm as a function of different stresses in (111) Si channel.

Fig. 7 The carrier concentration effective mass ($m_c$) of the three lowest valence subbands at gate electric field 1 MV/cm as a function of compressive and tensile stresses in (110) Si channel.

Fig. 8 The carrier concentration effective mass ($m_c$) of the three lowest valence subbands at gate electric field 1 MV/cm as a function of compressive and tensile stresses in (111) Si channel.

Fig. 9 The conductivity effective mass ($m_\sigma$) of the three lowest valence subbands at gate electric field 1 MV/cm as a function of compressive and tensile stresses in (110) Si channel.

Fig. 10 The conductivity effective mass ($m_\sigma$) of the three lowest valence subbands at gate electric field 1 MV/cm as a function of compressive and tensile stresses in (111) Si channel.

Fig. 11 The mobility in parallel and perpendicular directions relative to channel direction at gate electric field 1 MV/cm as a function of compressive and tensile stresses in (110) and (111) Si channels, respectively.

Fig. 1

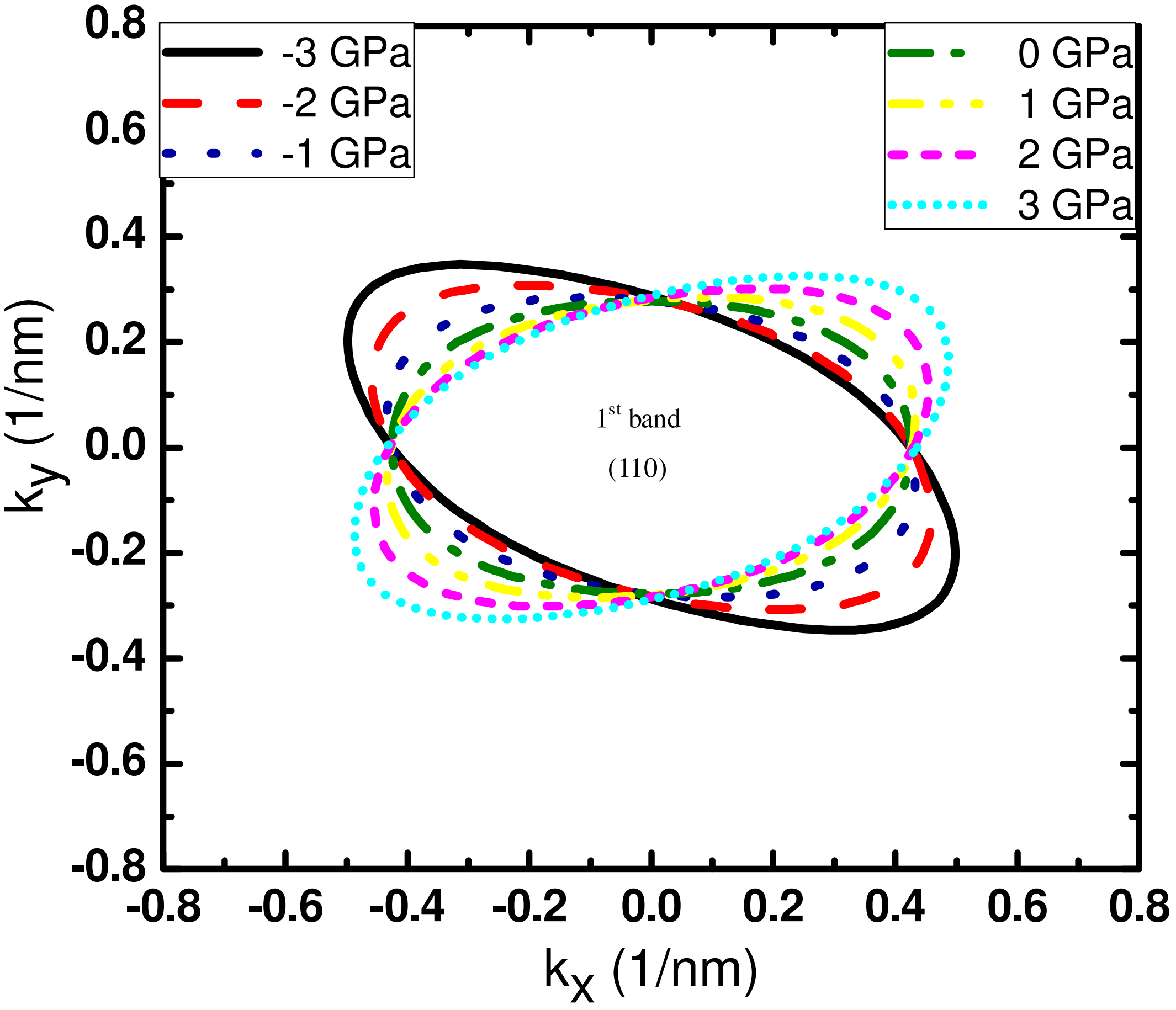

Fig. 2

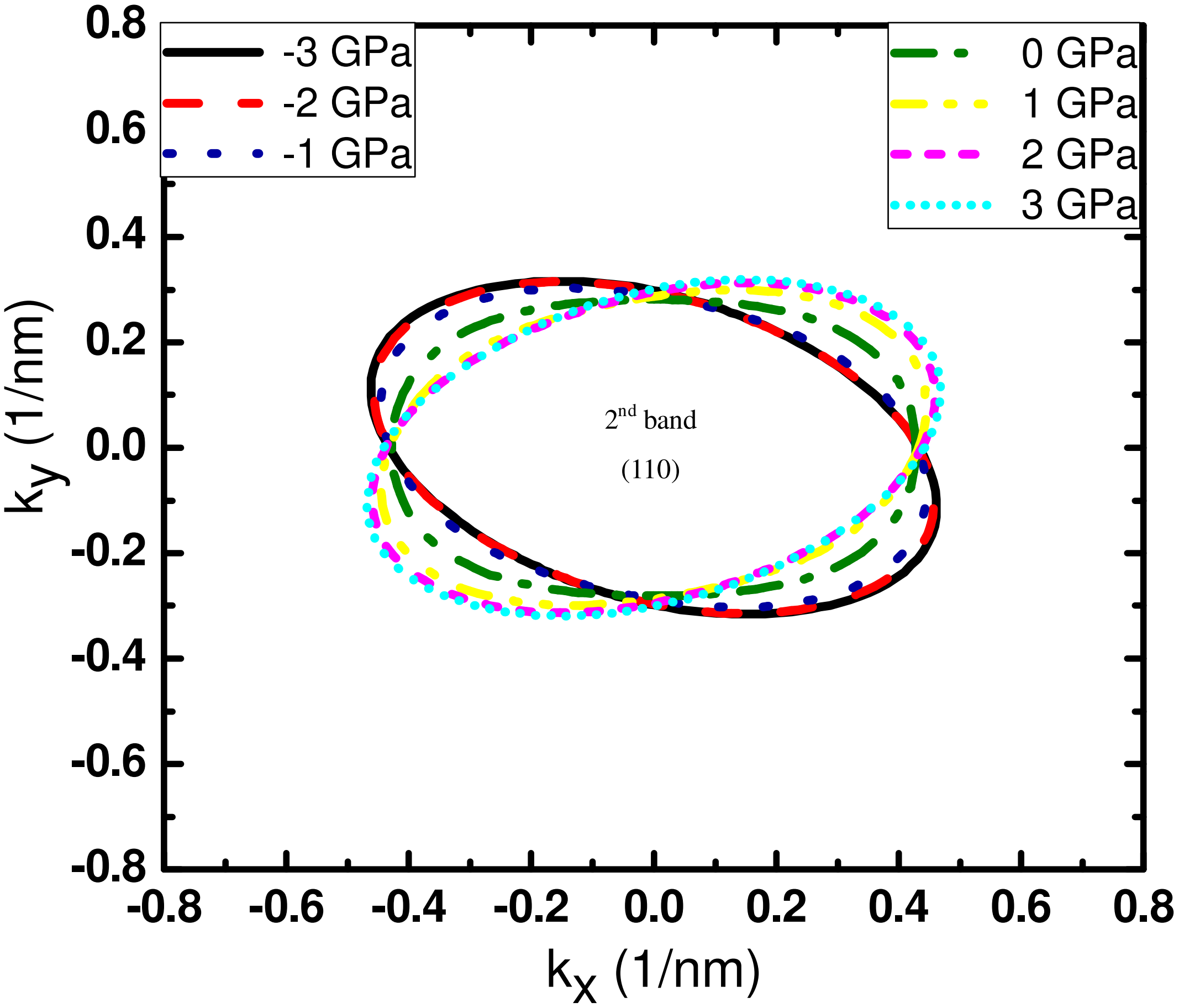

Fig. 3

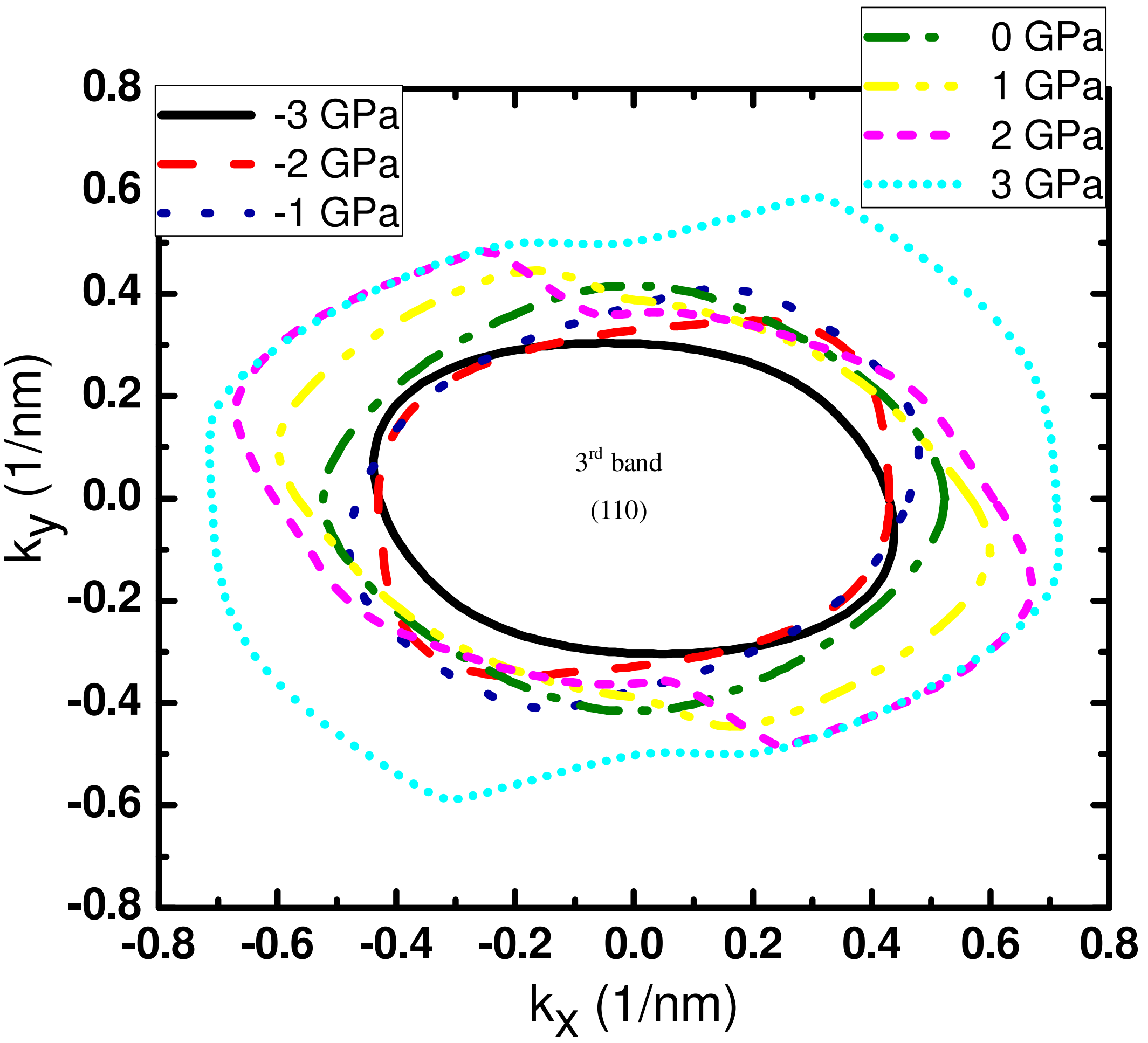

Fig. 4

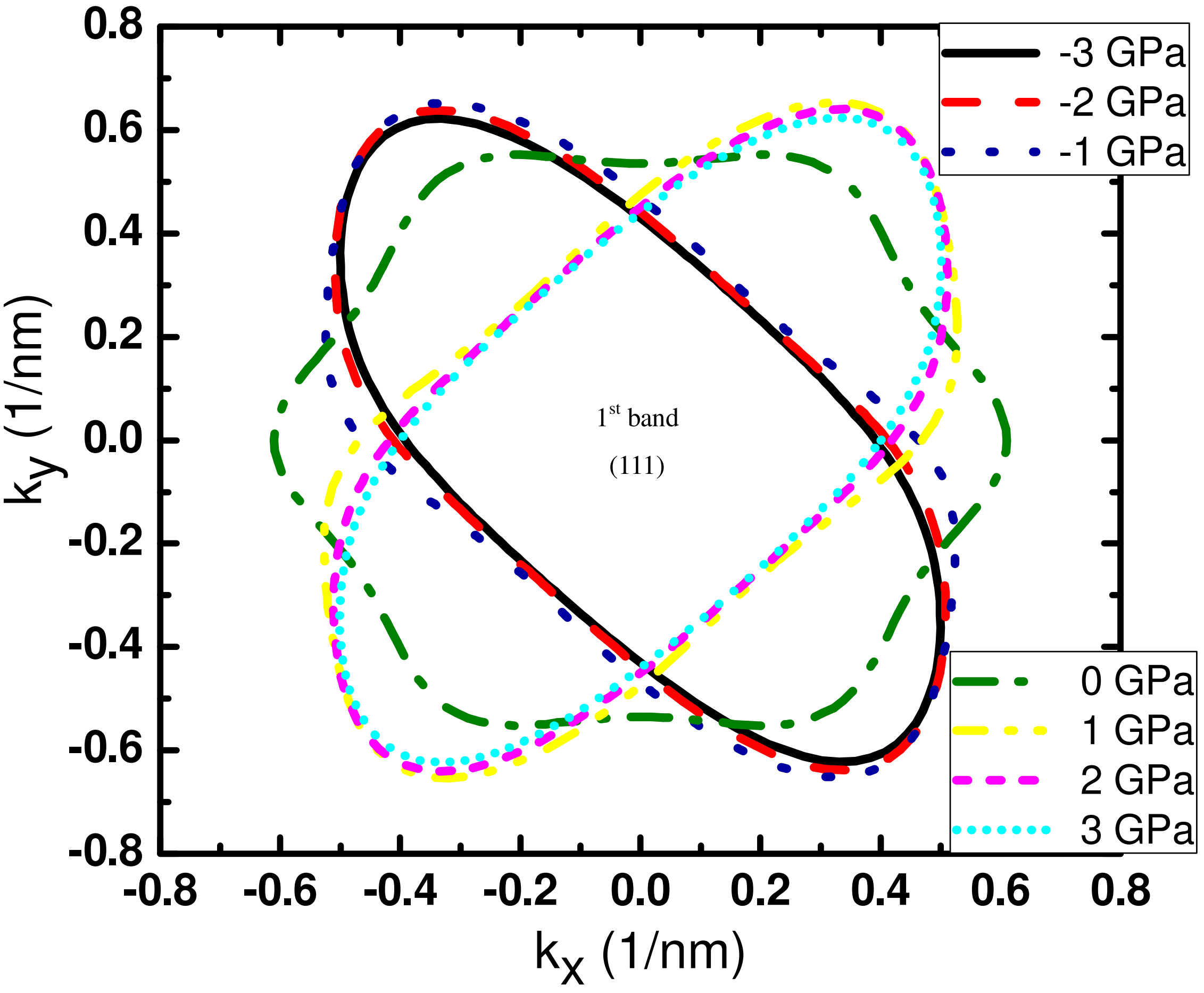

Fig. 5

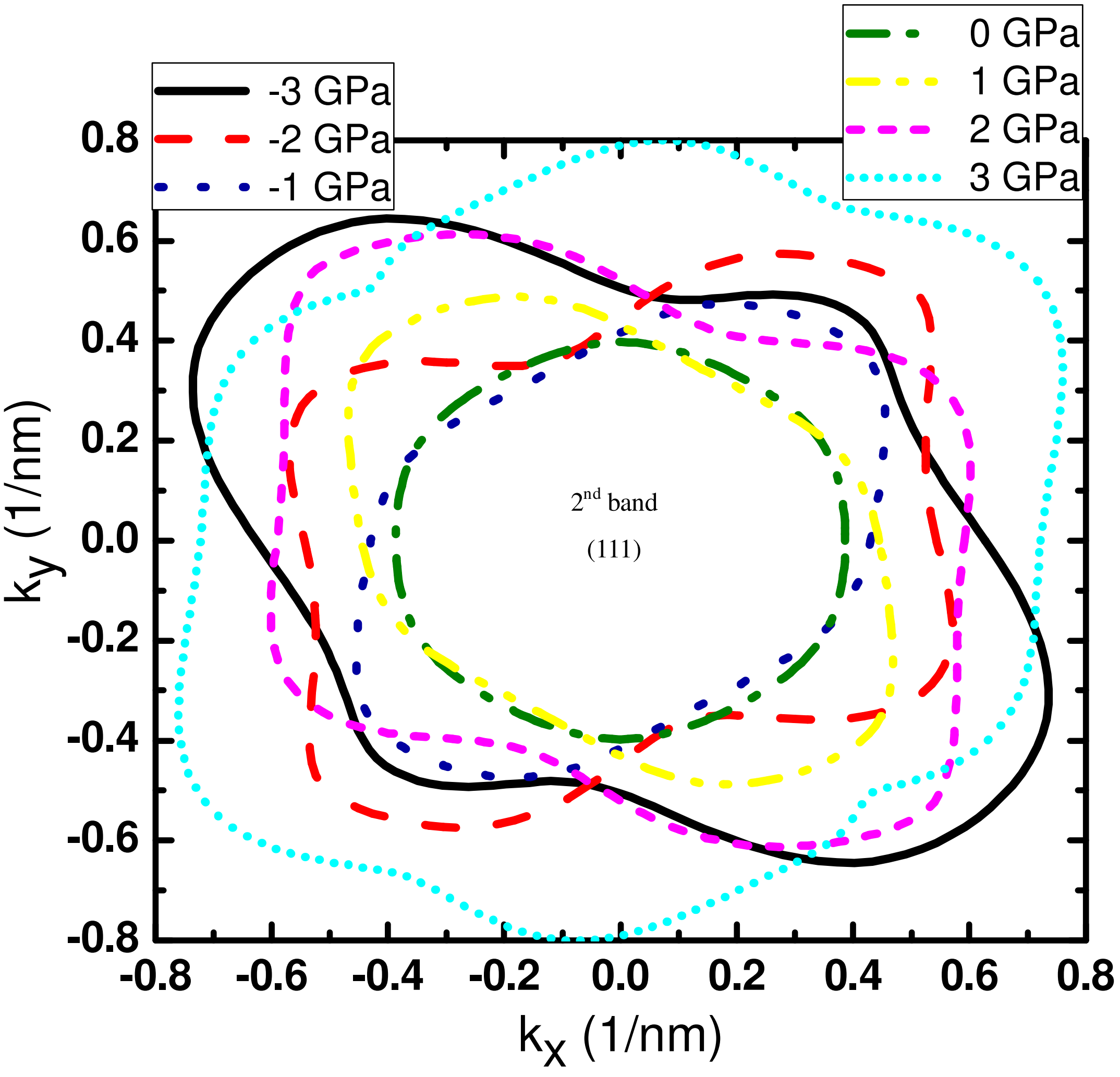

Fig. 6

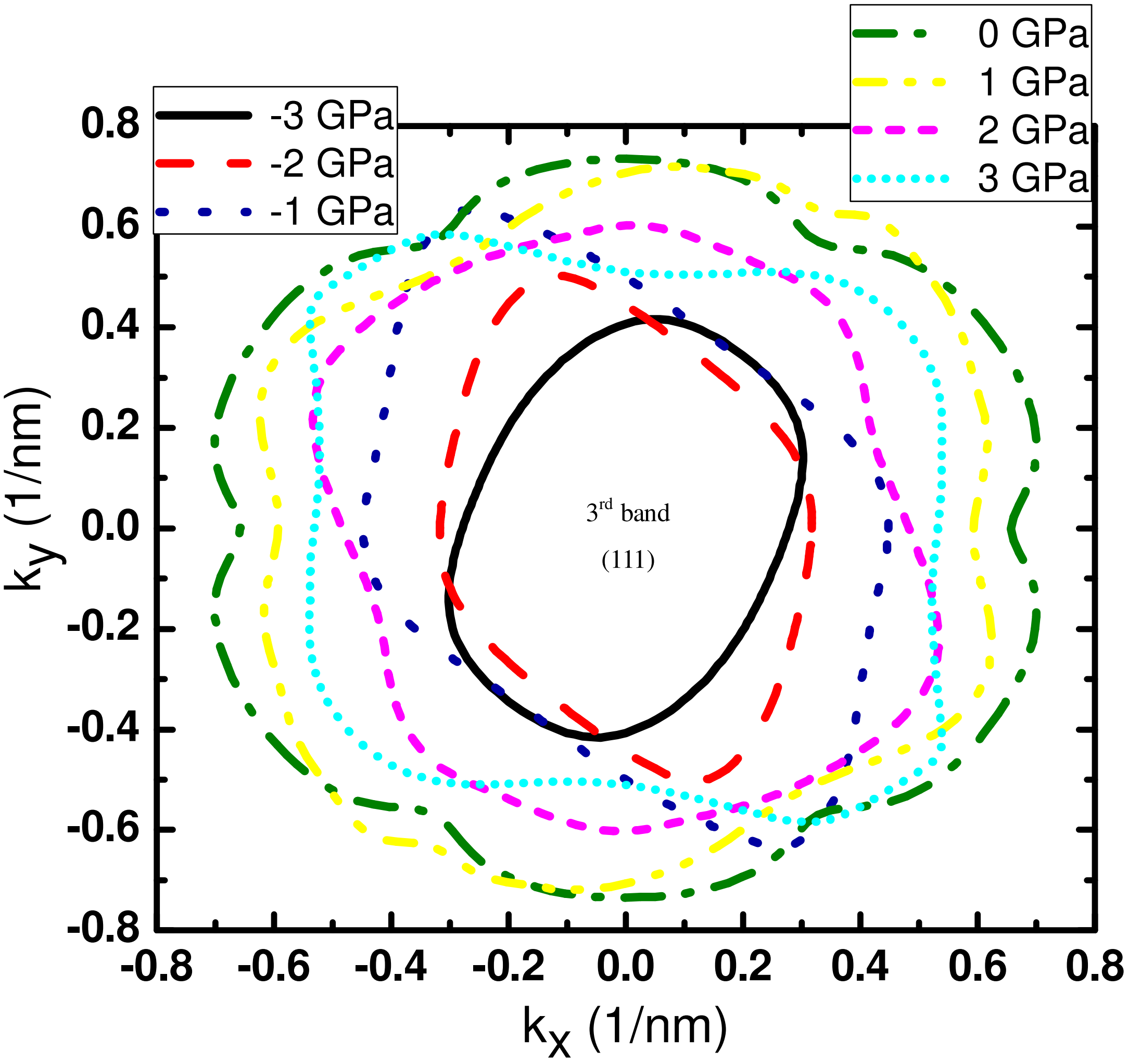

Fig. 7

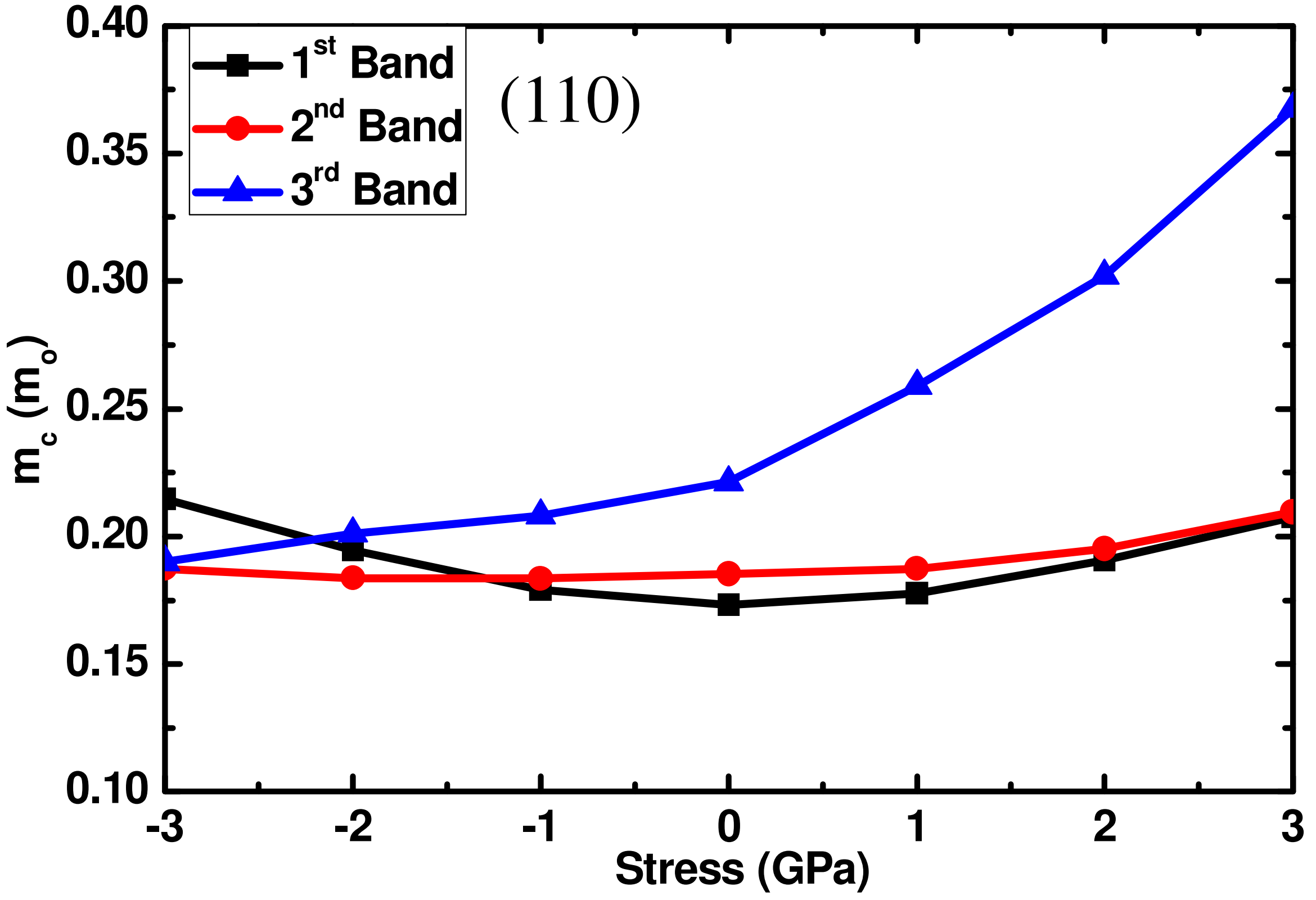

Fig. 8

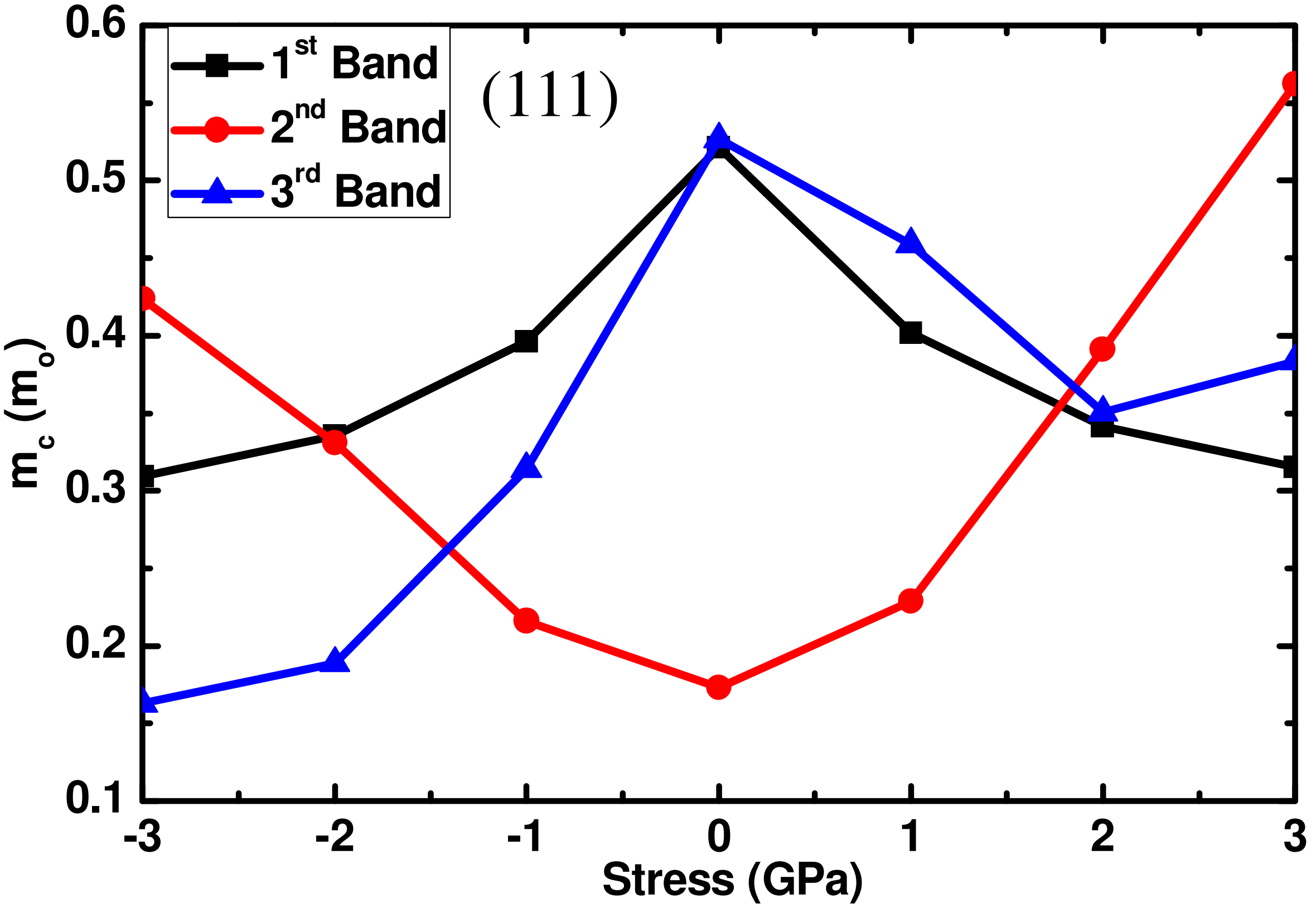

Fig. 9

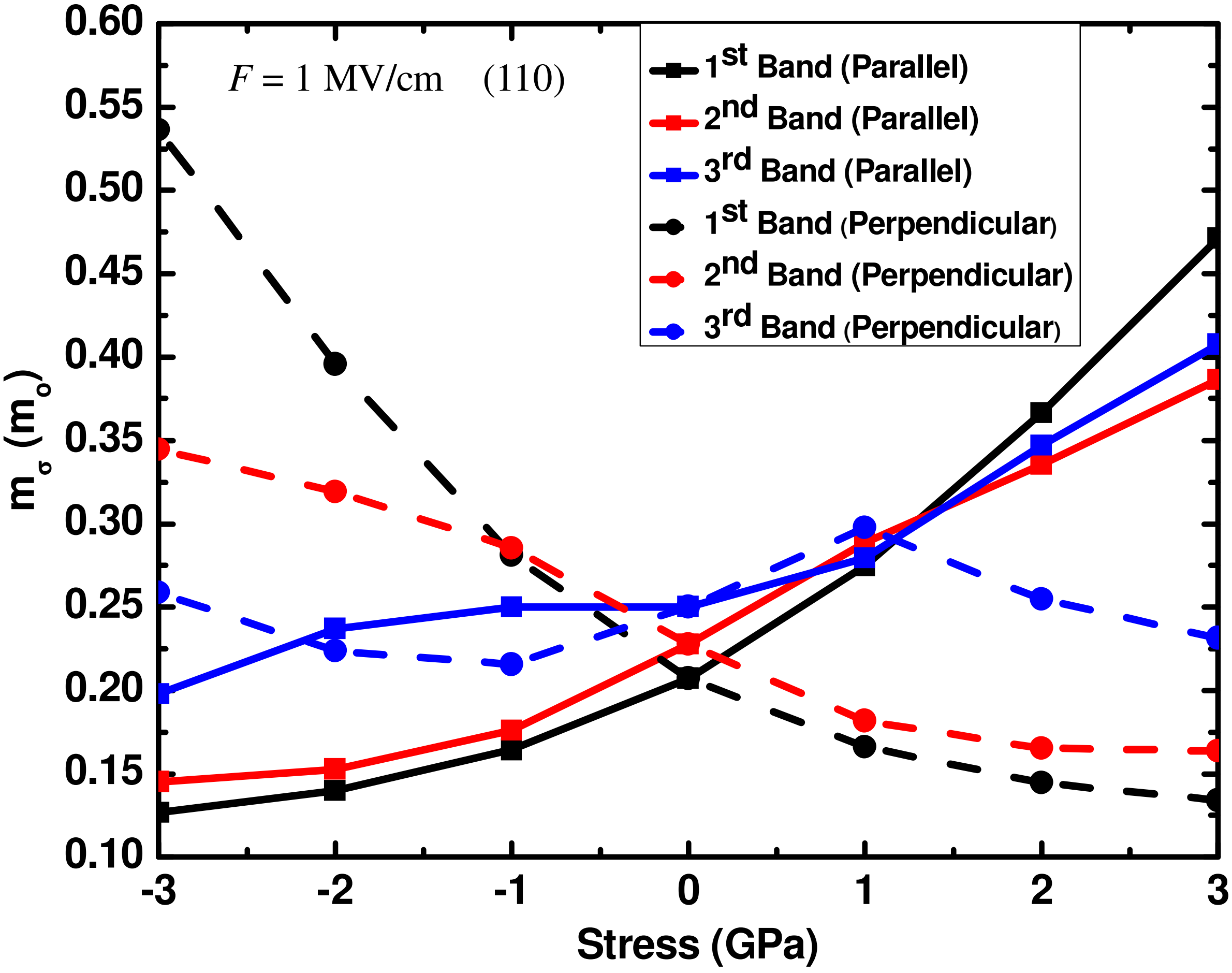

Fig. 10

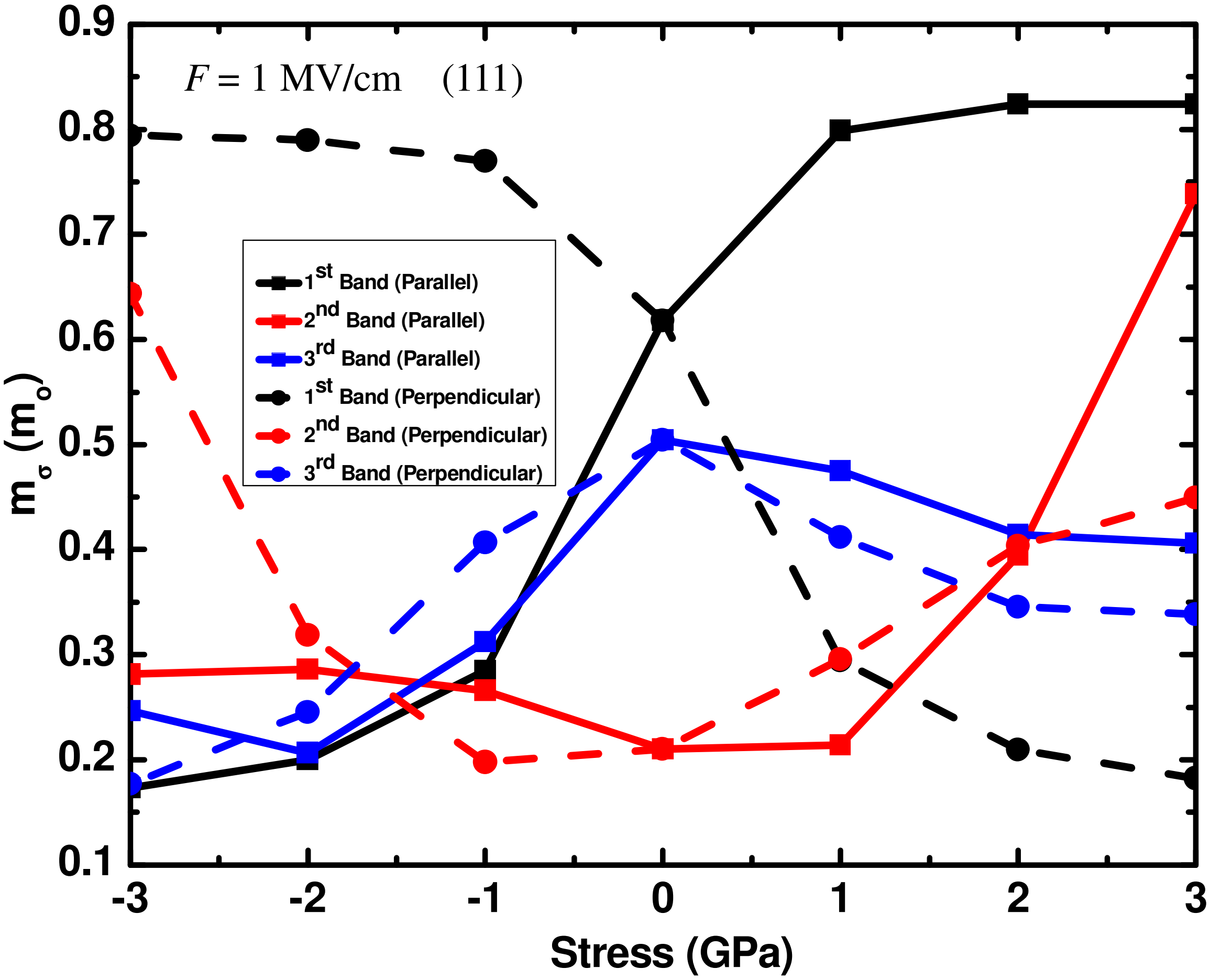

Fig. 11

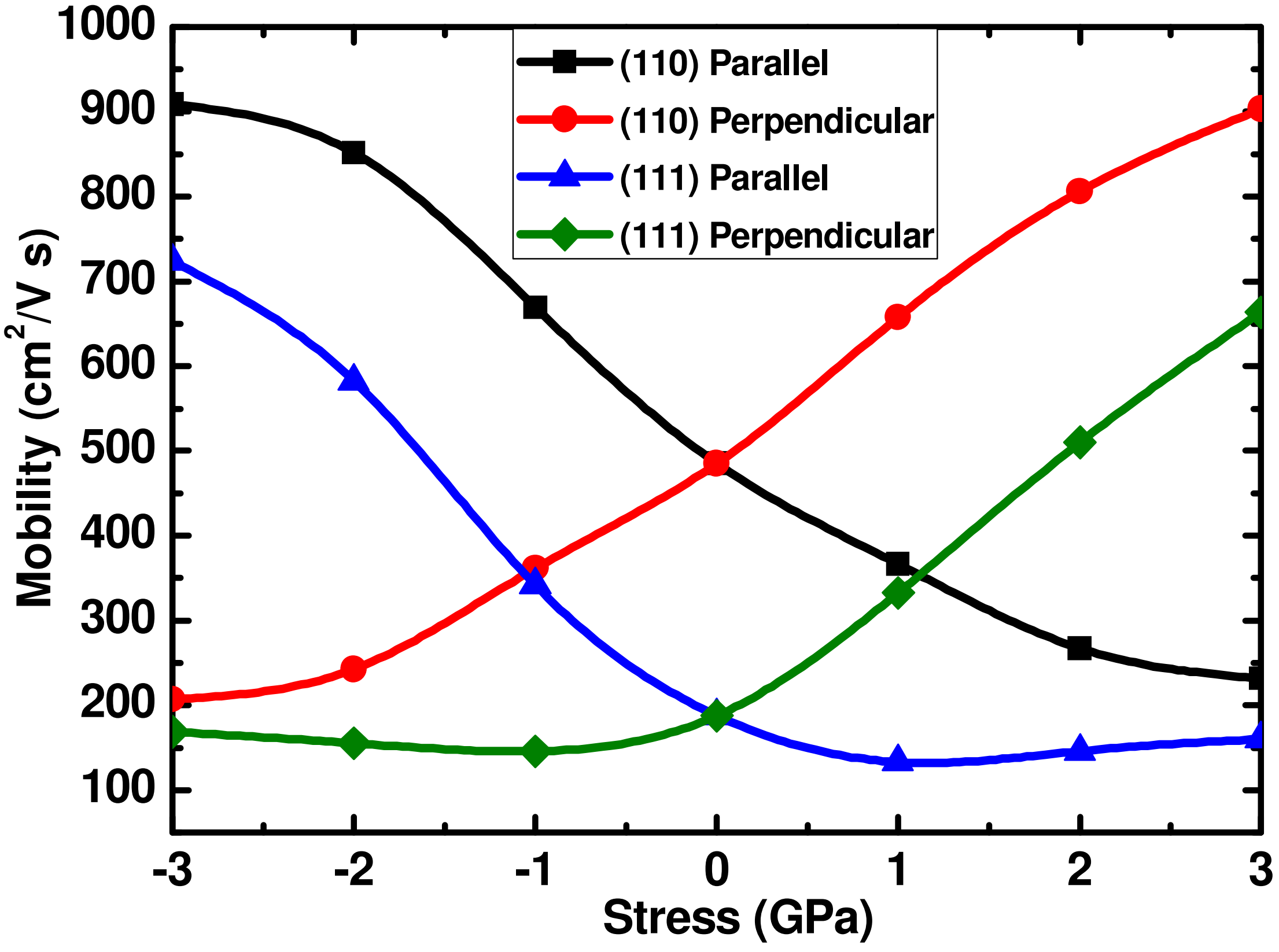